\documentclass{article}
\usepackage{booktabs, longtable}
\usepackage[english]{babel}
\usepackage{todonotes}
\usepackage[letterpaper,top=2cm,bottom=2cm,left=3cm,right=3cm,marginparwidth=1.75cm]{geometry}

\usepackage{amsmath}
\usepackage{graphicx}
\usepackage{paralist}
\usepackage[colorlinks=true, allcolors=blue]{hyperref}

\title{SAT, MaxSAT, and SMT for
QLDPC Distance Computation:\\
A Large-Scale Empirical Study}
\author{
Yu-Fang Chen$^{1}$ \quad
Seyed Mohammad Reza Jafari$^{1,2}$ \quad
Ching-Yi Lai$^{3}$ \\[0.5em]
$^{1}$Academia Sinica \\
$^{2}$National Taiwan University\\
$^{3}$National Yang Ming Chiao Tung University
}

\begin{document}
\maketitle

\begin{abstract}
Exact distance computation for quantum LDPC (QLDPC) codes plays a central role in validating candidate fault-tolerant quantum-code constructions, yet the computational structure of this problem remains poorly understood. Despite substantial recent progress in QLDPC design, it remains unclear which algorithmic principles govern the practical scalability of exact distance computation and which classes of exact solvers are best suited to this task.

To address these questions, we conduct a systematic study of SAT- and MaxSAT-based formulations for exact QLDPC distance computation across representative bivariate and generalized bicycle, lifted product, and quantum Tanner codes. We further compare these formulations against several established exact-distance approaches, including mixed-integer programming, Brouwer--Zimmermann-style search, and connected-cluster methods, in order to better understand the algorithmic landscape of exact QLDPC distance computation.

Our study challenges and refines several prevailing intuitions about exact QLDPC distance computation. First, despite the XOR-rich structure of QLDPC parity checks, practical scalability appears to be governed more by the handling of cardinality constraints and optimization bounds than by parity reasoning alone. Accordingly, XOR-aware reasoning does not provide a systematic advantage across our benchmark suite. 

Second, Brouwer-Zimmermann-style search, long regarded as the benchmark paradigm for exact distance computation in sparse classical codes, no longer maintains its traditional scalability advantage in the QLDPC setting. This finding challenges the expectation that techniques successful for sparse classical codes remain dominant for QLDPC codes.

Third, substantial qualitative differences arise even among MaxSAT solvers themselves. Branch-and-bound MaxSAT significantly outperforms unsat-core-based MaxSAT on challenging benchmarks, demonstrating that solver architecture and optimization strategy play a decisive role in practical scalability.

Our experimental platform is available as an open-source project at \url{https://github.com/guluchen/QDistSAT}.
\end{abstract}

\section{Introduction}
Modern quantum low-density parity-check (QLDPC) code constructions, including quantum Tanner codes, bivariate and generalized bicycle codes, and lifted product codes, have emerged as some of the most promising approaches toward scalable fault-tolerant quantum computation. These constructions achieve sparse parity-check structure together with increasingly strong distance and rate properties, making them among the most actively studied families of quantum error-correcting codes.

A fundamental challenge in the study of QLDPC codes is \emph{exact distance computation}. The code distance determines the minimum-weight logical operator and directly characterizes the worst-case error-correcting capability of the code. Unlike parameters such as the number of physical qubits or sparsity, the distance is often difficult to determine directly from the code construction itself, especially for large modern QLDPC codes. As modern QLDPC constructions continue to scale in size and complexity, exact distance computation has increasingly become a major computational bottleneck in practical code design and analysis.

A variety of computational approaches have been proposed for exact distance computation, including connected cluster methods, mixed-integer programming (MIP), and SAT-based formulations. Among these, SAT-based approaches are particularly appealing because they allow quantum distance computation to leverage decades of advances in SAT solving. Modern SAT solvers have achieved remarkable success on highly structured industrial verification and optimization problems. Yet, despite the growing importance of QLDPC codes, their performance on quantum distance computation instances has received comparatively little systematic study. This raises a natural question: which solver techniques are most effective for the unique constraint structures arising from QLDPC distance computation? By translating low-weight logical operators into Boolean constraints, SAT solvers can either construct explicit logical operators or certify their nonexistence below a given weight threshold. However, the resulting formulas differ substantially from conventional SAT benchmarks. QLDPC distance computation naturally generates parity constraints, logical-operator constraints, and weight-bounded search conditions, giving rise to highly structured optimization landscapes whose impact on solver performance remains poorly understood.  To the best of our knowledge, no prior work has systematically compared SAT-, SMT-, and MaxSAT-based approaches on a common collection of QLDPC distance computation benchmarks.

In this paper, we present a large-scale empirical study of SAT-, MaxSAT-, and SMT-based approaches for exact QLDPC distance computation across several major families of QLDPC codes, including bivarate and generalized bicycle codes, lifted product codes, and quantum Tanner codes. Our evaluation compares modern CDCL SAT solvers, XOR-aware SAT solvers, SMT solvers, and MaxSAT solvers under multiple cardinality constraint formulation schemes and constraint-handling strategies.

Our experiments reveal several surprising trends. MaxSAT approaches consistently outperform conventional SAT and SMT workflows on difficult QLDPC benchmarks and additionally outperform several non-SAT exact-distance approaches, including MIP-based and algebraic methods. Among all evaluated configurations, MaxCDCL exhibits the strongest robustness and successfully solves several challenging instances that remain intractable for competing approaches even under long timeout limits. In contrast, native XOR-aware reasoning provides only limited advantages despite the parity-rich structure of QLDPC constraints. Finally, cardinality-constraint formulations have a first-order impact on scalability, often producing performance differences comparable to those between solver families.

Overall, our results suggest that the dominant computational challenge in exact QLDPC distance computation lies not in parity reasoning alone, but in the interaction between cardinality constraints and proof of optimality. These findings reveal substantial qualitative differences among modern SAT-, SMT-, and MaxSAT-based approaches and provide new insight into the algorithmic structure of exact QLDPC distance computation.

\section{QLDPC Codes and Exact Distance Computation}

QLDPC codes form an important class of quantum error-correcting codes characterized by sparse parity-check structure and strong coding performance. A CSS-type~\cite{CS96,Ste96} QLDPC code is defined by two binary parity-check matrices \(H_X\) and \(H_Z\) satisfying the commutation condition
\[
H_X H_Z^T = 0.
\]

A central problem in the study of such codes is \emph{distance computation}. Intuitively, the code distance is the minimum number of qubits that must be affected to create an undetectable logical error.

For CSS codes, errors can be represented as binary vectors indicating which qubits are affected. Most errors violate one or more parity constraints induced by the code and are therefore detectable. However, some errors satisfy all parity constraints while still changing the encoded logical state. Such errors are called \emph{nontrivial logical operators}.

A \emph{valid nontrivial logical operator} must satisfy two key properties:
\begin{itemize}
    \item it must satisfy the \emph{parity constraints} induced by the code so that the error remains undetectable,
    \item it must be \emph{nontrivial}, meaning that it cannot be obtained from linear combinations of the parity-check rows themselves,
\end{itemize}
The code distance is then defined by the minimum possible \emph{weight} among all such nontrivial logical operators, corresponding to the smallest number of affected qubits.

More formally, the \(Z\)-distance \(d_Z\) is defined as the minimum-weight binary vector \(z\) satisfying
\[
H_X z^T = 0
\]
while not belonging to the row space of \(H_Z\). The first condition enforces the parity constraints induced by the code, while the second ensures nontriviality.

Symmetrically, the \(X\)-distance \(d_X\) is defined as the minimum-weight binary vector \(x\) satisfying
\[
H_Z x^T = 0
\]
while not belonging to the row space of \(H_X\).
The overall code distance is then given by
\[
d = \min(d_X,d_Z).
\]

\section{SAT and MaxSAT Formulations for Distance Computation}

Boolean satisfiability (SAT) is the problem of determining whether there exists an assignment to a collection of Boolean variables satisfying a given set of logical constraints. In SAT, constraints are typically represented in \emph{conjunctive normal form} (CNF), where a formula consists of a conjunction of \emph{clauses}, and each clause is a disjunction ($\lor$) of Boolean variables (e.g, $x_i$) or their negations (e.g, $\neg x_i$).
For example, consider the CNF formula
\[
(x_1 \lor x_2)
\;\land\;
(\neg x_1 \lor x_3)
\;\land\;
(x_2 \lor \neg x_3).
\]

A SAT solver attempts to determine whether there exists an assignment of truth values to the variables such that all clauses are simultaneously satisfied.

In this example, the assignment
\[
(x_1,x_2,x_3) = (0,1,0)
\]
satisfies all three constraints. Therefore, the instance is satisfiable.

Modern SAT solvers are highly optimized tools for solving large combinatorial search problems and have been successfully applied in areas such as hardware verification, software analysis, cryptography, and automated reasoning.

Maximum satisfiability (MaxSAT)\footnote{More precisely, the version considered here is the more general \emph{partial MaxSAT}, which allows both hard and soft clauses. Classical MaxSAT consists only of soft clauses.} extends SAT from feasibility to optimization. In MaxSAT, constraints are divided into two categories:
\begin{itemize}
    \item \emph{hard clauses}, which must always be satisfied, and
    \item \emph{soft clauses}, which the solver attempts to satisfy as many as possible.
\end{itemize}

For example, suppose we require the hard clause
\[
x_1 \lor x_2,
\]
together with the soft clauses
\[
\neg x_1,\qquad \neg x_2,\qquad \neg x_3.
\]

Since the hard clause requires at least one of \(x_1\) or \(x_2\) to be true, not all soft clauses can be satisfied simultaneously. The assignment
\[
(x_1,x_2,x_3) = (1,0,0)
\]
satisfies two soft clauses, namely \(\neg x_2\) and \(\neg x_3\), while
\[
(1,1,0)
\]
satisfies only one soft clause. A MaxSAT solver therefore prefers the first assignment because it satisfies more soft clauses.

Intuitively, SAT asks whether a \emph{feasible solution} exists, while MaxSAT searches for the 
\emph{best feasible solution} according to an optimization objective.

\subsection{SAT Formulation for Distance Computation}

Given a CSS code specified by parity-check matrices \(H_X\) and \(H_Z\), we introduce a Boolean variable \(x_i\) for each physical qubit \(i\). Intuitively,
\[
x_i = 1
\]
indicates that the candidate logical operator acts on qubit \(i\).

The SAT formulation consists of three major components corresponding to the three conditions (parity, nontriviality, and weight constraints) discussed in the previous section.

\paragraph{Parity constraints}
First, the candidate operator must satisfy all \emph{parity constraints} induced by the code. For \(Z\)-distance computation, this requires
\[
H_X x^T = 0.
\]

Operationally, each row of \(H_X\) imposes an even-parity constraint on a subset of variables. For example, suppose a row of \(H_X\) acts on qubits \(1,2,4\). Then the corresponding constraint becomes
\[
x_1 \oplus x_2 \oplus x_4 = 0,
\]
meaning that an even number of these variables must be assigned to \(1\).
For instance, $(x_1,x_2,x_4) = (1,1,0)$ satisfies the constraint, while $(1,0,0)$ does not.

These parity constraints can be encoded either directly as XOR constraints, which are supported natively by some solvers, or translated into CNF using Tseitin-style transformations~\cite{tseitin1983complexity}.

\paragraph{Nontriviality constraints}
Second, the candidate operator must satisfy the \emph{nontriviality constraint}. For \(Z\)-distance computation, this requires
\[
x \notin \operatorname{row}(H_Z),
\]
ensuring that the candidate operator cannot be generated from linear combinations of the parity-check rows themselves. Directly encoding this non-membership condition in SAT is inconvenient, as it would require reasoning about all possible linear combinations of the rows of \(H_Z\).

Let \(L_X\) and \(L_Z\) be dual \(X\)- and \(Z\)-logical operator matrices of the CSS code defined by \((H_X,H_Z)\), satisfying
\[
L_XL_Z^T=I.
\] 
For a reduction to SAT, one can instead check the equivalent condition
\[
L_Xx^T\neq 0.
\]
Since vectors in \(\operatorname{row}(H_Z)\) commute with all \(X\)-logical operators, while every nontrivial \(Z\)-logical operator anticommutes with at least one row of \(L_X\), the above condition is equivalent to \(x\notin\operatorname{row}(H_Z)\).

For example, if
\[
L_X=
\begin{bmatrix}
1&0&0&1
\end{bmatrix},
\]

then

\[
L_Xx^T = x_1 \oplus x_4.
\]

Hence the nontriviality constraint \(L_Xx^T\neq 0\) is encoded simply as

\[
x_1 \oplus x_4.
\]

If instead

\[
L_X=
\begin{bmatrix}
1&0&1&0\\
0&0&0&1
\end{bmatrix},
\]

then we introduce

\[
y_1=x_1\oplus x_3,
\qquad
y_2=x_4,
\]

and impose

\[
y_1\vee y_2.
\]

\paragraph{Weight constraint}
Third, the candidate operator must satisfy a \emph{weight constraint}. To determine whether the code distance is at most some target value \(w\), we bound the number of qubits on which the operator acts. This can be expressed using the cardinality constraint
\[
\sum_i x_i \leq w.
\]

For example, if
\[
w = 2,
\]
then assignments such as
\[
(x_1,x_2,x_3,x_4) = (1,1,0,0)
\]
are allowed, while
\[
(1,1,1,0)
\]
violates the constraint because the operator acts on three qubits.

Unlike parity constraints, cardinality constraints admit many different CNF formulations, including \emph{sequential-counter}~\cite{sinz2005towards}, \emph{totalizer}-based~\cite{bailleux2003efficient}, and a \emph{binary encoding} used in~\cite{ehatamm2026end}. Since this is a benchmarking paper, we decided to omit the details on how to translate the cardinality constraints to CNF and refer interested readers to the reference papers.

The resulting SAT instance is satisfiable if and only if there exists a nontrivial logical operator of weight at most \(w\).
Exact distance computation can then be performed incrementally by solving the SAT formulation for increasing values of \(w\), starting from \(w=1\). If the SAT instance is unsatisfiable for some \(w\), then the code distance must be greater than \(w\). The smallest value of \(w\) for which the instance becomes satisfiable is therefore the exact code distance.

\subsection{MaxSAT Formulation for Distance Computation}

The SAT formulation determines whether there exists a nontrivial logical operator whose weight is at most a given bound \(w\). Exact distance computation is therefore performed by repeatedly solving SAT instances for increasing values of \(w\).

Maximum satisfiability (\emph{MaxSAT}) provides an alternative formulation that avoids this iterative threshold search. Instead of checking feasibility for a fixed weight bound, the problem is formulated directly as an optimization problem minimizing the weight of the logical operator.

As in the SAT formulation, the \emph{parity constraints} and \emph{nontriviality constraints} are encoded as hard clauses that must always be satisfied. To minimize the operator weight, each variable \(x_i\) is additionally associated with a soft clause
\[
(\neg x_i),
\]
encouraging the solver to assign \(x_i = 0\).

For example, suppose the candidate operator is represented by four variables
\[
x_1,x_2,x_3,x_4.
\]
Then the soft clauses become
\[
(\neg x_1)
\;\land\;
(\neg x_2)
\;\land\;
(\neg x_3)
\;\land\;
(\neg x_4).
\]
An assignment such as
\[
(1,0,0,0)
\]
satisfies three soft clauses, while
\[
(1,1,0,0)
\]
satisfies only two. The MaxSAT solver therefore prefers the first assignment because it corresponds to a lower-weight logical operator.
This formulation directly computes a minimum-weight nontrivial logical operator without requiring repeated SAT solving over multiple candidate distance bounds. 

\section{Experimental Methodology}

To evaluate the effectiveness of SAT- and MaxSAT-based distance computation, we conducted a comprehensive empirical study across multiple families of QLDPC CSS codes, SAT solving paradigms, and constraint formulation strategies. Our evaluation focuses on three representative QLDPC code families: quantum Tanner codes~\cite{leverrier2022quantum}, generalized bicycle codes~\cite{bravyi2024high}, and lifted product codes~\cite{panteleev2022asymptotically}.

All experiments were conducted on a Linux server running Ubuntu 24.04 LTS, equipped with two AMD EPYC 7742 64-core processors (256 hardware threads in total) and 2 TB of DDR4 main memory.
We use the exact distance computation formulations described in the previous section and compare modern SAT, MaxSAT, and SMT solvers under multiple constraint formulation strategies to evaluate their effectiveness and scalability on exact QLDPC distance computation problems. 

\subsection{Benchmark Families of QLDPC CSS Codes}
Our evaluation focuses on three representative families of QLDPC CSS codes: bivariate and generalized bicycle codes, lifted product codes, and quantum Tanner codes. These families were selected because they exhibit substantially different structural properties, sparsity patterns, and distance behaviors, leading to diverse SAT and MaxSAT instances for exact distance computation. 

Benchmark instances were collected from multiple sources, summarized in Table~\ref{tab:benchmark-codes}. The resulting benchmark suite spans a wide range of code lengths, parity-check structures, and target distances, providing a diverse collection of exact distance computation workloads across modern QLDPC constructions.

\paragraph{Bivariate and Generalized Bicycle Codes.}
Bivariate bicycle (BB) codes and generalized bicycle (GB) codes are closely related families of algebraically constructed QLDPC CSS codes derived from sparse circulant operators and commuting polynomial representations. They are among the most widely studied finite-length QLDPC codes due to their relatively simple construction and strong empirical performance. The resulting parity-check matrices are highly sparse and structured, producing SAT instances with strong algebraic regularity. 

\paragraph{Lifted Product Codes.}
Lifted product codes generalize hypergraph-product-style constructions by combining graph lifts with product constructions. These codes often achieve larger block lengths while maintaining sparse parity-check matrices. Compared to BB codes, lifted product codes typically exhibit more irregular graph structure and larger search spaces, making exact distance computation substantially more challenging. 

\paragraph{Quantum Tanner Codes.}
Quantum Tanner codes are constructed from high-dimensional expander-based Tanner constructions and have recently attracted significant attention due to their asymptotic properties and improved distance scaling behavior. Their parity-check structures are typically less algebraically regular than BB codes and may induce particularly difficult combinatorial search spaces for SAT-based distance computation. 

Together, these three families provide a diverse collection of QLDPC CSS instances spanning different algebraic constructions, graph structures, sparsity patterns, and distance regimes. 

\begin{table}[htbp]
  \centering
  \caption{Benchmark CSS codes (maximum row and column weights of $H_x$, $H_z$). Benchmarking naming follow the pattern \texttt{\{family\}\_\{n\}\_\{k\}\_\{d\}}; a question mark denotes minimum distance not yet certified (e.g.\ \texttt{BB\_288\_12\_?}). }
  \label{tab:benchmark-codes}
  \begin{tabular}{lcc}
    \toprule
    Name & Row wt & Col wt \\
    \midrule
    LP\_34\_20\_2~\cite{webster2026distance}      & 8 & 5 \\
    LP\_136\_32\_4~\cite{webster2026distance}     & 8 & 5 \\
    LP\_238\_44\_6~\cite{webster2026distance}     & 8 & 5 \\
    LP\_340\_56\_8~\cite{webster2026distance}     & 8 & 5 \\
    LP\_442\_68\_10~\cite{webster2026distance}    & 8 & 5 \\
    LP\_544\_80\_12~\cite{xu2024constant,webster2026distance}    & 8 & 5 \\
    LP\_714\_100\_?~\cite{xu2024constant,webster2026distance}    & 8 & 5 \\
    LP\_1768\_224\_?~\cite{webster2026distance}   & 8 & 5 \\
    BB\_108\_8\_10~\cite{bravyi2024high}     & 6 & 3 \\
    BB\_144\_12\_12~\cite{bravyi2024high}    & 6 & 3 \\
    BB\_144\_14\_14~\cite{wang2026check}    & 8 & 4 \\
    BB\_288\_12\_?~\cite{bravyi2024high}     & 6 & 3 \\
    BB\_360\_12\_?~\cite{webster2026distance}     & 6 & 3 \\
    BB\_72\_12\_6~\cite{bravyi2024high}      & 6 & 3 \\
    BB\_90\_8\_10~\cite{bravyi2024high}      & 6 & 3 \\
    GB\_144\_12\_8~\cite{11154497}     & 7 & 5 \\
    GB\_144\_12\_12~\cite{11154497}    & 8 & 4 \\
    TN\_200\_10\_10~\cite{Radebold2025DistanceFinding}    & 12 & 9 \\
    TN\_250\_10\_15~\cite{Radebold2025DistanceFinding}    & 12 & 8 \\
    TN\_36\_8\_3~\cite{Radebold2025DistanceFinding}       & 6 & 4 \\
    TN\_54\_11\_4~\cite{Radebold2025DistanceFinding}      & 6 & 4 \\
    TN\_72\_14\_4~\cite{Radebold2025DistanceFinding}      & 6 & 4 \\
    TN\_108\_2\_12~\cite{webster2026distance}     & 9 & 8 \\
    TN\_144\_2\_13~\cite{webster2026distance}     & 9 & 8 \\
    TN\_360\_4\_?~\cite{webster2026distance}      & 9 & 8 \\
    TN\_36\_8\_4~\cite{webster2026distance}       & 9 & 8 \\
    TN\_72\_8\_8~\cite{webster2026distance}       & 9 & 8 \\
    \bottomrule
  \end{tabular}
\end{table}

\subsection{Solvers}

Our evaluation includes a diverse collection of SAT, SMT, and MaxSAT solvers representing several major solving paradigms used in modern Boolean constraint solving.

\paragraph{SAT solvers.}
The evaluated SAT solvers primarily belong to the conflict-driven clause learning (CDCL) family and span a range of branching heuristics, restart policies, and clause-management strategies. In our implementation, these solvers are accessed uniformly through a lightweight abstraction layer built on top of PySAT’s \texttt{Solver} interface, including clause insertion, satisfiability testing, and model extraction. We tried all solvers supported by PySAT, including Minisat(v2.2)~\cite{sorensson2005minisat}, Glucose(v3, v4, v4.2)~\cite{audemard2018glucose}, CaDiCal(v1.95, v1.53, v1.03)~\cite{biere2024cadical}, Lingeling~\cite{biere2013lingeling}, MapleSat~\cite{liang2016learning}, and Megesat(v3)~\cite{manthey2021mergesat}.

\subparagraph{XOR constraints:} Recall that the SAT encoding contains three types of constraints: parity, nontriviality, and weight.
Both parity and nontriviality constraints heavily use the XOR operators.
Those standard CDCL solvers handle the XOR constraints through Tseitin-style CNF translation~\cite{tseitin1983complexity}. We also evaluated
CryptoMiniSat~\cite{soos2016cryptominisat} that supports native XOR reasoning.
\subparagraph{Cardinality reasoning:} 
Recall that cardinality constraints admit a variety of CNF encodings, including the \emph{sequential counter}~\cite{sinz2005towards}, \emph{totalizer}-based encodings~\cite{bailleux2003efficient}, and the \emph{binary encoding} adopted in~\cite{ehatamm2026end}. We evaluated all of these encodings using standard CDCL solvers. In addition, we experimented with solvers that provide native support for cardinality constraints, including MiniCard~\cite{liffiton2012cardinality} and Gluecard (v3 and v4)~\cite{ignatiev2018pysat}.

\paragraph{SMT solvers.}
We additionally evaluated the SMT solvers Z3~\cite{de2008z3} and CVC5~\cite{barbosa2022cvc5}. Unlike the SAT backends. Both solvers support higher-level Boolean and arithmetic reasoning beyond CNF formulas, including native XOR reasoning and richer cardinality constraint handling. They were included to evaluate whether such higher-level reasoning improves exact QLDPC distance computation compared to purely CNF-based CDCL workflows.

\paragraph{MaxSAT solvers.}
Most modern MaxSAT solvers are \emph{unsat-core-guided}, including Open-WBO~\cite{martins2014open}, EvalMaxSAT~\cite{avellaneda2023evalmaxsat}, and CASHWMaxSAT~\cite{pan2024efficient}, all of which are top-performing solvers in the MaxSAT Evaluation competition. We also evaluate the RC2 family~\cite{ignatiev2019rc2}, which supports multiple underlying SAT backends, including Glucose, CaDiCaL, CryptoMiniSat, and MiniSat22. In contrast to unsat-core-guided approaches, MaxCDCL~\cite{li2021combining} employs a branch-and-bound framework integrated with CDCL and local-core reasoning. 

To facilitate systematic experimentation with exact QLDPC distance computation, we developed \textsc{DistQLDPC}~\cite{chen_distqldpc}, an open-source framework derived from MaxCDCL and specialized for distance-solving tasks. Besides supporting the original MaxCDCL workflow, \textsc{DistQLDPC} provides a unified experimental interface for evaluating multiple cardinality-constraint encodings, including sequential-counter, totalizer, and related formulations. Importantly, when executed with its default configuration, \textsc{DistQLDPC} reproduces the behavior and performance of the original MaxCDCL implementation.

A distinguishing feature of \textsc{DistQLDPC} is its support for incremental certification. Exact distance computation can be viewed as a sequence of feasibility queries that test the existence of logical operators below a target weight threshold. Consequently, even when the full computation does not converge, SAT-based formulations can still certify nontrivial lower bounds on the code distance. To expose analogous intermediate information for optimization-oriented solvers, \textsc{DistQLDPC} explicitly tracks and reports the best certified lower and upper bounds available at termination. 

\subparagraph{SAT formulations expose incremental certification behavior.}

Although branch-and-bound MaxSAT and mixed-integer programming generally exhibit stronger robustness on hard benchmark instances, SAT-based formulations retain a distinct advantage in incremental certification. By iteratively testing the existence of logical operators below a target weight threshold, SAT-based solving can still establish certified lower bounds on the code distance even when the full exact-distance computation times out. In contrast, optimization-oriented approaches primarily focus on locating an optimal solution and typically do not provide comparable intermediate certificates before convergence. 

To better expose this intermediate behavior in optimization-oriented solving, we implement the experimental framework \textsc{DistQLDPC}~\cite{chen_distqldpc}, which extends MaxCDCL with additional instrumentation for exact-distance computation. In particular, \textsc{DistQLDPC} explicitly reports the current certified lower and upper bounds when the computation terminates before convergence. The framework additionally supports multiple cardinality-constraint formulations beyond the default MaxCDCL setting, enabling controlled comparisons across sequential-counter, totalizer, and related formulations within a unified interface. Importantly, when executed using the default configuration, the performance of \textsc{DistQLDPC} is identical to the original MaxCDCL implementation.

\paragraph{Non-SAT solvers.}
As baselines, we compare against several exact distance computation methods implemented in the \texttt{codeDistance} framework~\cite{webster2026distance}. These include \textsc{GurobiDist} and \textsc{MIPDist}, which formulate distance computation as mixed-integer programming problems; the Magma implementation of the Brouwer--Zimmermann algorithm; and \textsc{dist\_m4ri\_CC}, which implements a connected-cluster-based search algorithm.

\begin{table}[t]
\centering
\caption{Summary of evaluated solvers and their solving paradigms.}
\label{tab:solver-types}
\small
\begin{tabular}{ll}
\toprule
Type & Solvers \\
\midrule
CDCL SAT
& Minisat, Glucose, CaDiCal, Lingeling, MapleSAT, MergeSat \\
SAT with native XOR reasoning
& CryptoMiniSat \\
SAT with native cardinality reasoning
& MiniCard, Gluecard \\
SMT
& Z3, CVC5 \\
Unsat-core-guided MaxSAT
& Open-WBO, EvalMaxSAT, CASHWMaxSAT, RC2-family \\
Branch-and-bound MaxSAT
& MaxCDCL, DistQLDPC \\
Non-SAT baselines
& GurobiDist, MIPDist, Magma, Dist\_M4RI\_CC \\
\bottomrule
\end{tabular}
\end{table}

\subsection{Evaluation}

Our evaluation involves a large design space spanning benchmark instances, solver backends, and encoding strategies. Table~\ref{tab:benchmark-codes} contains 27 benchmark instances, while Table~\ref{tab:solver-types} includes 21 exact-distance computation tools covering SAT, MaxSAT, SMT, and non-SAT-based approaches. Many of these tools additionally admit multiple configurations, primarily differing in the formulation of cardinality constraints. Performing a full-scale evaluation over all solver--configuration combinations under long timeout limits would require substantial computational resources and would also obscure the effect of encoding choices themselves. We therefore begin with a focused pre-evaluation study to identify representative and competitive configurations for the main experiments.

\paragraph{Cardinality-constraint formulation pre-evaluation and experiment design.}

Cardinality constraints admit several standard CNF formulations, including the \emph{sequential counter}~\cite{sinz2005towards}, \emph{totalizer}~\cite{bailleux2003efficient}, and \emph{binary} formulations used in prior work~\cite{ehatamm2026end}. Because cardinality constraints constitute a central component of our SAT approach, the choice of formulation can have a substantial impact on propagation strength, clause structure, and ultimately solver scalability.

To evaluate this effect, we compare all 11 SAT solvers using these three cardinality-constraint formulations on the 27 benchmark instances, with a timeout limit of 3 minutes per run. Across the resulting $11 \times 27 = 297$ solver-benchmark combinations (Table~\ref{tab:card-encoding}) per cardinality constraint formulation, 225 successfully compute the exact distance under at least one formulation. Among these cases, the sequential-counter formulation achieves the best runtime in 113 cases, while the totalizer formulation achieves the best runtime in 92 cases. In contrast, we do not observe any benchmark instance in which the binary-based formulation attains the best performance.

These results indicate that the choice of cardinality constraint formulation has a first-order effect on solver scalability, often comparable to the choice of solver backend itself. In particular, the binary-based formulation appears consistently dominated by the sequential-counter and totalizer formulations on our benchmark suite. We therefore restrict the remaining SAT-based evaluations to the sequential-counter and totalizer formulations. This removes the empirically dominated binary-based formulation and reduces the SAT-based configuration space by approximately 20\%, while preserving the competitive constraint formulation strategies identified in the pre-evaluation.

\begin{table}[t]
  \centering
  \caption{Cardinality constraint formulation: aggregate results (891 runs, 180\,s timeout).}
  \label{tab:card-encoding}
  \begin{tabular}{lrrrr}
    \toprule
    Formulation & Solved & Timeout & Fastest \\
    \midrule
    \texttt{seqcounter}  & 218 &  79 & 133 \\
    \texttt{totalizer} & 224 &  73 &  92 \\
    \texttt{binary}      & 151 & 146 &   0 \\
    \bottomrule
  \end{tabular}
\end{table}

\begin{figure*}[p]
    \centering

    \includegraphics[
        width=1\textwidth
    ]{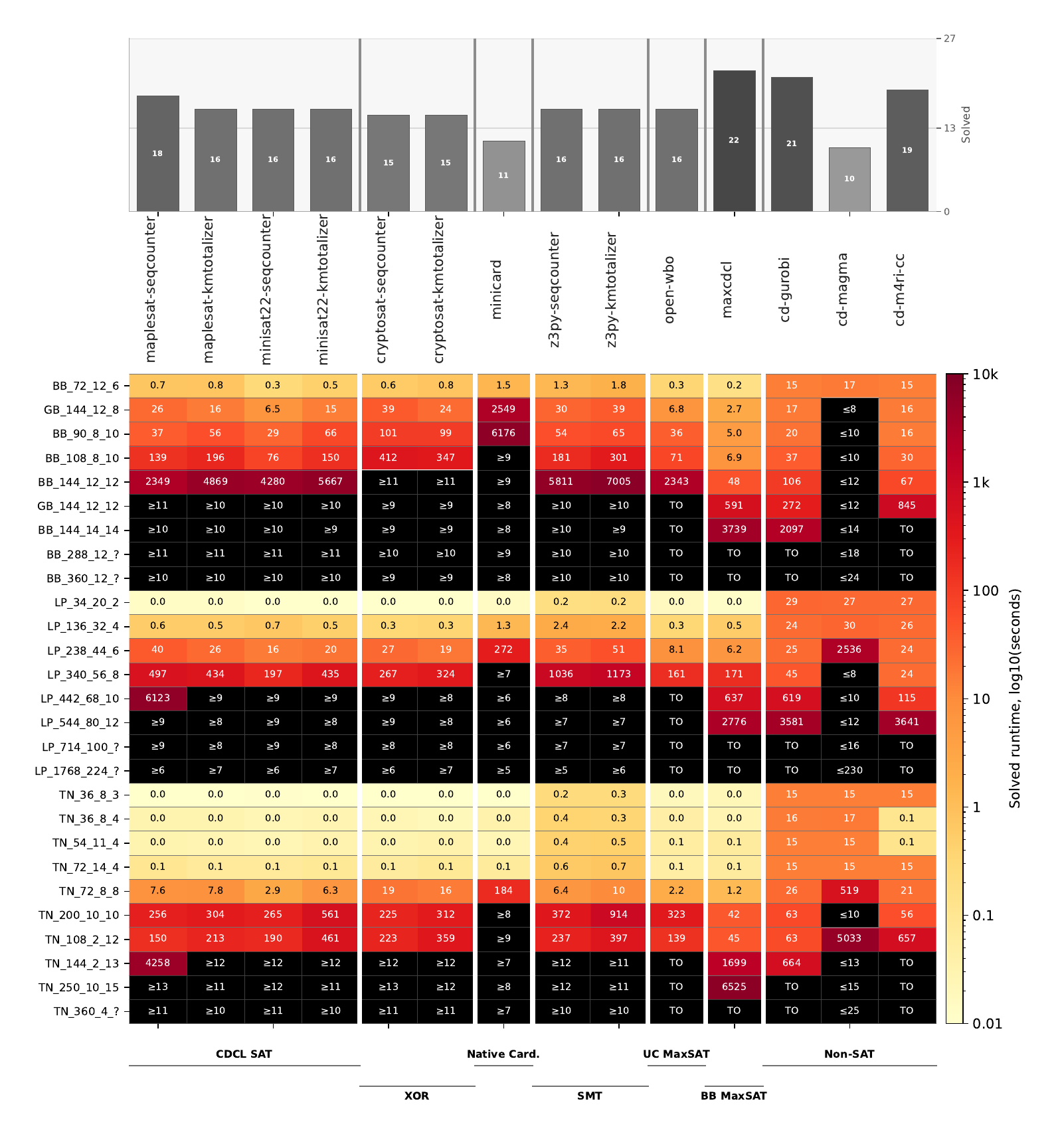}\\

    \caption{
    Solver-behavior heatmap.
    Light colors indicate faster solved runtimes, while black cells denote timeout runs.
    Solver configurations are grouped by solving paradigm. Cell: run time in seconds. Black Cell: timeout with obtained distance upper/lower bounds. Bar chart: number of solved cases within 2 hours.
    }
    \label{fig:curated-heatmap}
\end{figure*}

\paragraph{Large-scale solver evaluation.}
We evaluate the 27 benchmark instances from Table~\ref{tab:benchmark-codes} using all solvers listed in Table~\ref{tab:solver-types}, together with both sequential counter and totalizer cardinality constraint formulations whenever applicable. This results in a total of 46 solver configurations. Each solver-benchmark run is assigned a timeout limit of 2 hours.

To present the main experimental trends more clearly, Figure~\ref{fig:curated-heatmap} shows a curated subset of representative configurations, including the two strongest-performing CDCL SAT solvers; the strongest configurations with XOR support, native cardinality support, unsat-core-based MaxSAT, and branch-and-bound-style MaxSAT; and the strongest non-SAT exact solvers from each major category, including mixed-integer programming, clustering-based methods, and Brouwer-Zimmermann-based search. The complete experimental logs and the full 46-configuration heatmap are provided in the appendix. The full heatmap (Figure~\ref{fig:full-heatmap}) further shows that performance differences among variants within the same solver family are generally much smaller than the differences observed across solver paradigms and constraint formulation strategies, supporting our use of the strongest-performing representative configurations for the main comparison.

\paragraph{Results.}
Figure~\ref{fig:curated-heatmap} summarizes the performance of representative solver configurations across all benchmark instances. Rows correspond to benchmark codes, while columns correspond to solver configurations grouped by solving paradigm. Each cell reports the runtime (in seconds) required to compute the exact distance; lighter colors indicate faster runtimes. Black cells denote timeout runs, in which the displayed value represents the best certified lower or upper bound obtained before termination. The bar chart above the heatmap reports the total number of benchmark instances solved within the 2-hour timeout limit.

The figure reveals a sharp separation between optimization-oriented approaches and other exact-distance computation paradigms on hard QLDPC benchmark instances. While most solver configurations successfully compute the distance on easy and moderate benchmarks, the transition-region instances, such as \texttt{BB\_144\_12\_12}, \texttt{BB\_144\_14\_14}, \texttt{LP\_442\_68\_10}, \texttt{LP\_544\_80\_12}, \texttt{TN\_144\_2\_13}, and \texttt{TN\_250\_10\_15}, expose substantial differences in scalability and robustness. In this regime, branch-and-bound MaxSAT and mixed-integer-programming-based solving remain consistently competitive, whereas many SAT-, SMT-, and Brouwer-Zimmermann-based configurations begin to time out.

Among all evaluated configurations, MaxCDCL achieves the strongest overall coverage, solving one additional benchmark instance beyond the MIP-based Gurobi baseline within the 2-hour timeout limit. However, the advantage of MaxCDCL over Gurobi is not uniform. On several difficult benchmark instances, Gurobi achieves runtimes comparable to or faster than MaxCDCL itself.

\subparagraph{XOR-aware SAT solving does not provide a systematic advantage.} Despite the XOR-rich structure of both the parity-check constraints and the nontriviality constraints, XOR-aware SAT solving does not exhibit a clear systematic advantage on our benchmark suite. Across both easy and transition-region benchmark instances, the CryptoMiniSat-based configurations generally remain comparable to, or weaker than, conventional CDCL SAT configurations using the same cardinality-constraint formulation. In particular, XOR-aware reasoning does not significantly improve robustness on the harder benchmark instances where many SAT-based approaches begin to time out. These results suggest that the dominant computational difficulty in exact QLDPC distance computation is driven less by parity reasoning itself and more by the handling of cardinality constraints.

\subparagraph{Native cardinality support is less effective.}
Although native cardinality reasoning is designed specifically for handling pseudo-Boolean and cardinality constraints, the evaluated native-cardinality solvers, including \texttt{MiniCard}, \texttt{Gluecard3}, and \texttt{Gluecard4}, generally scale less effectively than SAT configurations using explicit sequential-counter or totalizer formulations. This gap becomes particularly pronounced on transition-region benchmark instances, where native-cardinality configurations frequently time out substantially earlier than conventional CDCL SAT solvers using explicit CNF formulations. These results suggest that, in the context of exact QLDPC distance computation, explicitly encoded cardinality constraints interact more effectively with modern clause-learning and propagation mechanisms than native cardinality reasoning itself.

\subparagraph{Classical-code distance techniques do not transfer cleanly to QLDPC benchmarks.} Among the evaluated non-SAT exact methods, the behavior of the Brouwer-Zimmermann-based Magma implementation is particularly notable. Brouwer-Zimmermann-style search is widely regarded as one of the strongest exact-distance techniques for classical codes~\cite{webster2026distance}, yet its performance degrades substantially on the evaluated QLDPC benchmark instances. While the method remains effective on many easy benchmark instances, it frequently times out in the transition region where optimization-oriented approaches, particularly MaxCDCL and Gurobi, continue to scale more effectively. These results suggest that exact QLDPC distance computation exhibits structural challenges not adequately addressed by classical-code distance techniques alone.

At the same time, many timeout runs produced by the Magma implementation already identify upper bounds matching the best exact distances eventually obtained by other methods. This indicates that Brouwer-Zimmermann-style search is often effective at locating low-weight candidate logical operators, but substantially weaker at proving that no better solution exists.

\subparagraph{Sequential-counter and totalizer formulations exhibit distinct solver interactions.}

The experimental results show that the formulation of cardinality constraints substantially affects exact QLDPC distance computation. Although both the sequential-counter and totalizer formulations remain competitive across the benchmark suite, their relative effectiveness depends strongly on the solver backend and benchmark structure. For example, on \texttt{BB\_144\_12\_12}, the MapleSAT sequential-counter configuration is approximately twice as fast as its totalizer counterpart, while on \texttt{LP\_340\_56\_8} the two formulations exhibit comparable performance under the same backend. Similarly, the MiniSat22 sequential-counter configuration substantially outperforms the corresponding totalizer formulation on several transition-region benchmark instances, whereas the relative difference is smaller for MapleSAT and CryptoMiniSat configurations.

These results indicate that the impact of cardinality-constraint formulations is highly solver-dependent and cannot be explained solely by the size or compactness of the resulting CNF formulas.

\subparagraph{Transition-region benchmarks are the most informative.}

The experimental results show that the most informative benchmark instances are those in the transition region, where some solver configurations successfully compute the exact distance while others time out or exhibit substantial runtime degradation. In contrast, uniformly easy instances provide limited discriminative power, while uniformly unsolved instances reveal little beyond the current scalability frontier.

Based on the current experimental landscape, we recommend the following benchmark instances as representative evaluation benchmarks for modern exact QLDPC distance solvers. For bivariate and generalized bicycle codes: \texttt{BB\_144\_12\_12}, \texttt{GB\_144\_12\_12}, and \texttt{BB\_144\_14\_14}. For lifted product codes: \texttt{LP\_442\_68\_10} and \texttt{LP\_544\_80\_12}. For quantum Tanner codes: \texttt{TN\_144\_2\_13} and \texttt{TN\_250\_10\_15}. These benchmark instances expose substantial differences in robustness across solving paradigms, cardinality-constraint formulations, and backend architectures, making them particularly suitable for evaluating modern exact-distance solvers.

Overall, the experimental results suggest that the dominant computational difficulty in modern exact QLDPC distance computation arises less from parity reasoning itself and more from the interaction between optimization structure, cardinality constraints, and backend search behavior. Across the evaluated benchmark suite, optimization-oriented approaches consistently define the strongest practical scalability frontier, while specialized XOR-aware reasoning alone does not yield a comparable advantage. At the same time, the experiments also show that formulation choices for cardinality constraints remain a first-order factor in solver scalability, often producing substantial runtime differences even under the same solving paradigm.

\begin{figure}[t]
\label{fig:cactus}
\centering
\includegraphics[width=\linewidth]{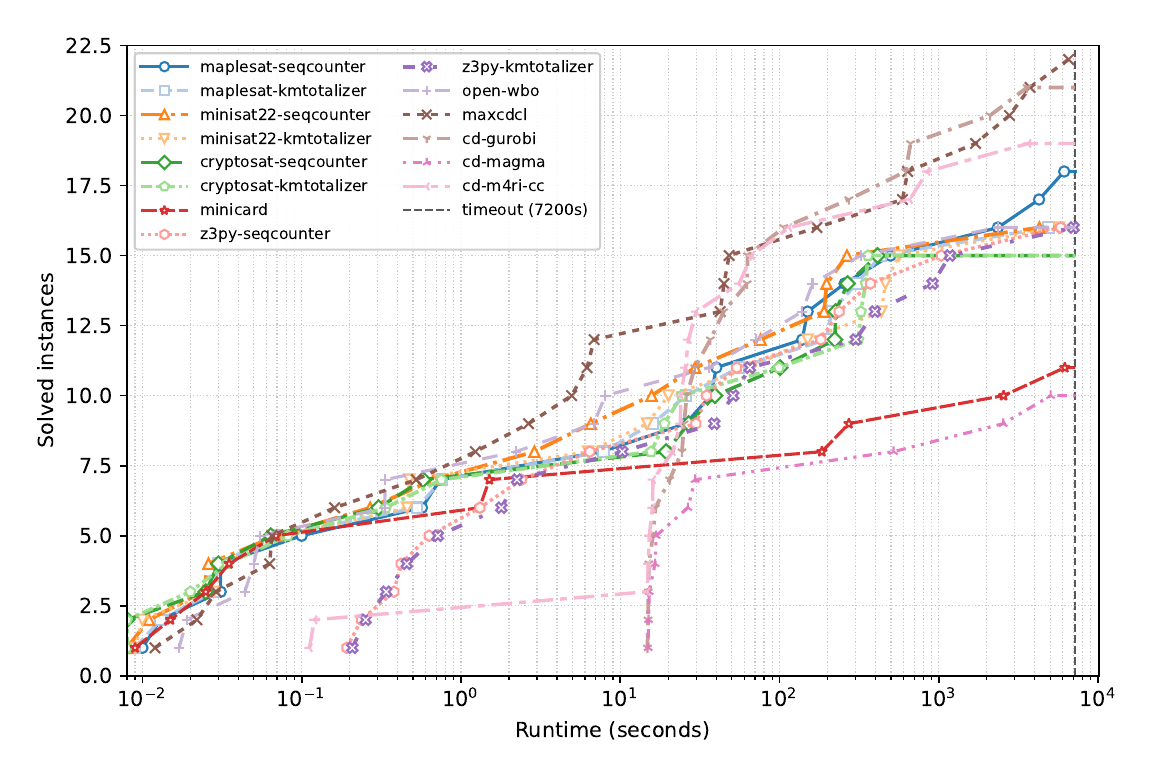}
\caption{
Solved-instances-versus-runtime plot for the curated representative solver configurations.
Each curve shows how many benchmark instances are solved within a given runtime threshold.
Curves extending farther to the right indicate stronger solved-instance coverage within the 2-hour timeout limit.
}
\end{figure}

\paragraph{Cumulative solved-count over time.} Figure~\ref{fig:cactus} presents a solved-instances-versus-runtime plot for the curated representative solver configurations used in Figure~\ref{fig:curated-heatmap}. For each configuration, the curve shows the number of benchmark instances solved within a given runtime threshold. This view complements the heatmap by emphasizing overall scalability and timeout-frontier behavior rather than detailed per-benchmark runtime comparisons.

Figure~\ref{fig:cactus} suggests a rough stratification across solver paradigms. Branch-and-bound MaxSAT, mixed-integer programming, and clustering-based exact search define the strongest scalability frontier, continuing to solve transition-region benchmark instances after most other paradigms plateau. Conventional CDCL SAT, XOR-aware SAT, and unsat-core-based MaxSAT remain competitive on moderate benchmark instances but generally reach the timeout frontier earlier on the hardest cases. In contrast, native-cardinality SAT and the Brouwer--Zimmermann-based Magma implementation exhibit substantially earlier scalability collapse on the evaluated benchmark suite.

\section{Concluding Remarks}

Recent work by Webster, Jacob, and Higgott~\cite{webster2026distance} provided a broad survey of distance-finding methodologies across classical codes, CSS quantum codes, and non-CSS quantum codes, including both exact and heuristic approaches. In contrast, our work focuses specifically on exact QLDPC CSS distance computation, motivated by the central role of QLDPC codes in scalable fault-tolerant quantum computation and the growing need for reliable exact-distance evaluation of newly proposed QLDPC code constructions.

Our experiments reveal several computational phenomena beyond aggregate solved-instance rankings. Most notably, the practical scalability of SAT- and MaxSAT-based approaches depends strongly on the underlying optimization architecture and cardinality-constraint handling strategy. Our experiments suggest that the dominant computational difficulty in exact QLDPC distance computation does not arise primarily from parity constraints alone. Instead, the main bottleneck appears to be proving the nonexistence of logical operators below a given cardinality threshold.

Our experimental results also suggest a more nuanced picture than solver recommendations based solely on high-level algorithm categories. Prior benchmarking study~\cite{webster2026distance} recommends clustering-based methods such as m4riCC as the primary exact approach for QLDPC-style sparse Tanner-graph instances, while SAT-based methods are typically treated as a secondary category. In contrast, our experiments show that branch-and-bound-style MaxSAT and carefully engineered CDCL SAT configurations remain highly competitive across transition-region QLDPC benchmarks, substantially outperforming unsat-core-based MaxSAT formulations and several other SAT-oriented configurations. These results suggest that treating ``SAT'' or ``MaxSAT'' as a single algorithmic category may obscure important computational distinctions in exact QLDPC distance computation.

Finally, our study suggests that transition-region benchmark instances, rather than uniformly easy or uniformly unsolved benchmarks, provide substantially more informative evaluation settings for exact-distance solvers. In particular, these instances more clearly expose differences in optimization strategy, scalability behavior, and cardinality-constraint handling across solver paradigms. We hope that the experimental comparisons and empirical observations presented in this work help establish a more systematic foundation for future research on exact QLDPC distance computation.

\bibliographystyle{alpha}
\bibliography{reference}
\newpage
\appendix
\section{Detailed Experiment Results}
\vspace{-5mm}
\begin{longtable}{llrrrl p{0.22\textwidth}}
\caption{Benchmark solver results for \texttt{BB\_108\_8\_10} ($n=108$, scan $d \leq 20$, timeout 7200\,s per configuration).} \label{tab:bench-BB_108_8_10} \\
\toprule
Solver & Strategy & Time (s) & Vars & Clauses & Result & Status \\
\midrule
\endhead
\midrule
\multicolumn{7}{r}{\textit{Continued on next page}} \\
\midrule
\endfoot
\bottomrule
\endlastfoot
z3py & seqcounter & 181.484 & 1320 & 2498 & 10 & \texttt{OK} \\
z3py & kmtotalizer & 300.782 & 726 & 1524 & 10 & \texttt{OK} \\
cvc5 & seqcounter & 333.008 & 1320 & 2498 & 10 & \texttt{OK} \\
cvc5 & kmtotalizer & 1815.820 & 726 & 1524 & 10 & \texttt{OK} \\
minisat22 & seqcounter & 76.019 & 2110 & 5534 & 10 & \texttt{OK} \\
minisat22 & kmtotalizer & 150.152 & 1516 & 4560 & 10 & \texttt{OK} \\
glucose3 & seqcounter & 85.990 & 2110 & 5534 & 10 & \texttt{OK} \\
glucose3 & kmtotalizer & 156.029 & 1516 & 4560 & 10 & \texttt{OK} \\
\end{longtable}

\clearpage
\begin{longtable}{llrrrl p{0.22\textwidth}}
\caption{Benchmark solver results for \texttt{BB\_144\_12\_12} ($n=144$, scan $d \leq 20$, timeout 7200\,s per configuration).} \label{tab:bench-BB_144_12_12} \\
\toprule
Solver & Strategy & Time (s) & Vars & Clauses & Result & Status \\
\midrule
\endhead
\midrule
\multicolumn{7}{r}{\textit{Continued on next page}} \\
\midrule
\endfoot
\bottomrule
\endlastfoot
z3py & seqcounter & 5811.085 & 2040 & 3890 & 12 & \texttt{OK} \\
z3py & kmtotalizer & 7004.855 & 977 & 2163 & 12 & \texttt{OK} \\
cvc5 & seqcounter & 7200.000 & 1671 & 3158 & $\geq$10 & \texttt{timeout} \\
cvc5 & kmtotalizer & 7200.000 & 1010 & 2375 & $\geq$9 & \texttt{timeout} \\
minisat22 & seqcounter & 4279.646 & 3158 & 8194 & 12 & \texttt{OK} \\
minisat22 & kmtotalizer & 5666.731 & 2095 & 6467 & 12 & \texttt{OK} \\
glucose3 & seqcounter & 2276.715 & 3158 & 8194 & 12 & \texttt{OK} \\
glucose3 & kmtotalizer & 5724.550 & 2095 & 6467 & 12 & \texttt{OK} \\
\end{longtable}

\clearpage
\begin{longtable}{llrrrl p{0.22\textwidth}}
\caption{Benchmark solver results for \texttt{BB\_144\_14\_14} ($n=144$, scan $d \leq 20$, timeout 7200\,s per configuration).} \label{tab:bench-BB_144_14_14} \\
\toprule
Solver & Strategy & Time (s) & Vars & Clauses & Result & Status \\
\midrule
\endhead
\midrule
\multicolumn{7}{r}{\textit{Continued on next page}} \\
\midrule
\endfoot
\bottomrule
\endlastfoot
z3py & seqcounter & 7200.000 & 1675 & 3162 & $\geq$10 & \texttt{timeout} \\
z3py & kmtotalizer & 7200.000 & 1014 & 2379 & $\geq$9 & \texttt{timeout} \\
cvc5 & seqcounter & 7200.000 & 1419 & 2654 & $\geq$8 & \texttt{timeout} \\
cvc5 & kmtotalizer & 7200.000 & 983 & 2109 & $\geq$7 & \texttt{timeout} \\
minisat22 & seqcounter & 7200.000 & 3505 & 10310 & $\geq$10 & \texttt{timeout} \\
minisat22 & kmtotalizer & 7200.000 & 2844 & 9527 & $\geq$9 & \texttt{timeout} \\
glucose3 & seqcounter & 7200.000 & 3505 & 10310 & $\geq$10 & \texttt{timeout} \\
glucose3 & kmtotalizer & 7200.000 & 2844 & 9527 & $\geq$9 & \texttt{timeout} \\
\end{longtable}

\clearpage
\begin{longtable}{llrrrl p{0.22\textwidth}}
\caption{Benchmark solver results for \texttt{BB\_288\_12\_?} ($n=288$, scan $d \leq 20$, timeout 7200\,s per configuration).} \label{tab:bench-BB_288_12_unknown} \\
\toprule
Solver & Strategy & Time (s) & Vars & Clauses & Result & Status \\
\midrule
\endhead
\midrule
\multicolumn{7}{r}{\textit{Continued on next page}} \\
\midrule
\endfoot
\bottomrule
\endlastfoot
z3py & seqcounter & 7200.000 & 3399 & 6470 & $\geq$10 & \texttt{timeout} \\
z3py & kmtotalizer & 7200.000 & 1906 & 4085 & $\geq$10 & \texttt{timeout} \\
cvc5 & seqcounter & 7200.000 & 2855 & 5386 & $\geq$8 & \texttt{timeout} \\
cvc5 & kmtotalizer & 7200.000 & 1939 & 4216 & $\geq$8 & \texttt{timeout} \\
minisat22 & seqcounter & 7200.000 & 6024 & 16118 & $\geq$11 & \texttt{timeout} \\
minisat22 & kmtotalizer & 7200.000 & 4262 & 13196 & $\geq$11 & \texttt{timeout} \\
glucose3 & seqcounter & 7200.000 & 5755 & 15582 & $\geq$10 & \texttt{timeout} \\
glucose3 & kmtotalizer & 7200.000 & 4262 & 13196 & $\geq$11 & \texttt{timeout} \\
\end{longtable}

\clearpage
\begin{longtable}{llrrrl p{0.22\textwidth}}
\caption{Benchmark solver results for \texttt{BB\_360\_12\_?} ($n=360$, scan $d \leq 20$, timeout 7200\,s per configuration).} \label{tab:bench-BB_360_12_unknown} \\
\toprule
Solver & Strategy & Time (s) & Vars & Clauses & Result & Status \\
\midrule
\endhead
\midrule
\multicolumn{7}{r}{\textit{Continued on next page}} \\
\midrule
\endfoot
\bottomrule
\endlastfoot
z3py & seqcounter & 7200.000 & 4263 & 8126 & $\geq$10 & \texttt{timeout} \\
z3py & kmtotalizer & 7200.000 & 2450 & 5205 & $\geq$10 & \texttt{timeout} \\
cvc5 & seqcounter & 7200.000 & 3575 & 6754 & $\geq$8 & \texttt{timeout} \\
cvc5 & kmtotalizer & 7200.000 & 2411 & 5192 & $\geq$8 & \texttt{timeout} \\
minisat22 & seqcounter & 7200.000 & 7305 & 19910 & $\geq$10 & \texttt{timeout} \\
minisat22 & kmtotalizer & 7200.000 & 5492 & 16989 & $\geq$10 & \texttt{timeout} \\
glucose3 & seqcounter & 7200.000 & 7305 & 19910 & $\geq$10 & \texttt{timeout} \\
glucose3 & kmtotalizer & 7200.000 & 5492 & 16989 & $\geq$10 & \texttt{timeout} \\
\end{longtable}

\clearpage
\begin{longtable}{llrrrl p{0.22\textwidth}}
\caption{Benchmark solver results for \texttt{BB\_72\_12\_6} ($n=72$, scan $d \leq 20$, timeout 7200\,s per configuration).} \label{tab:bench-BB_72_12_6} \\
\toprule
Solver & Strategy & Time (s) & Vars & Clauses & Result & Status \\
\midrule
\endhead
\midrule
\multicolumn{7}{r}{\textit{Continued on next page}} \\
\midrule
\endfoot
\bottomrule
\endlastfoot
z3py & seqcounter & 1.314 & 636 & 1166 & 6 & \texttt{OK} \\
z3py & kmtotalizer & 1.793 & 499 & 1049 & 6 & \texttt{OK} \\
cvc5 & seqcounter & 0.722 & 636 & 1166 & 6 & \texttt{OK} \\
cvc5 & kmtotalizer & 0.975 & 499 & 1049 & 6 & \texttt{OK} \\
minisat22 & seqcounter & 0.269 & 1216 & 3390 & 6 & \texttt{OK} \\
minisat22 & kmtotalizer & 0.459 & 1079 & 3273 & 6 & \texttt{OK} \\
glucose3 & seqcounter & 0.495 & 1216 & 3390 & 6 & \texttt{OK} \\
glucose3 & kmtotalizer & 0.408 & 1079 & 3273 & 6 & \texttt{OK} \\
\end{longtable}

\clearpage
\begin{longtable}{llrrrl p{0.22\textwidth}}
\caption{Benchmark solver results for \texttt{BB\_90\_8\_10} ($n=90$, scan $d \leq 20$, timeout 7200\,s per configuration).} \label{tab:bench-BB_90_8_10} \\
\toprule
Solver & Strategy & Time (s) & Vars & Clauses & Result & Status \\
\midrule
\endhead
\midrule
\multicolumn{7}{r}{\textit{Continued on next page}} \\
\midrule
\endfoot
\bottomrule
\endlastfoot
z3py & seqcounter & 53.800 & 1086 & 2048 & 10 & \texttt{OK} \\
z3py & kmtotalizer & 65.128 & 618 & 1284 & 10 & \texttt{OK} \\
cvc5 & seqcounter & 147.897 & 1086 & 2048 & 10 & \texttt{OK} \\
cvc5 & kmtotalizer & 561.606 & 618 & 1284 & 10 & \texttt{OK} \\
minisat22 & seqcounter & 29.493 & 1740 & 4558 & 10 & \texttt{OK} \\
minisat22 & kmtotalizer & 66.291 & 1272 & 3794 & 10 & \texttt{OK} \\
glucose3 & seqcounter & 23.397 & 1740 & 4558 & 10 & \texttt{OK} \\
glucose3 & kmtotalizer & 32.254 & 1272 & 3794 & 10 & \texttt{OK} \\
\end{longtable}

\clearpage
\begin{longtable}{llrrrl p{0.22\textwidth}}
\caption{Benchmark solver results for \texttt{GB\_144\_12\_12} ($n=144$, scan $d \leq 20$, timeout 7200\,s per configuration).} \label{tab:bench-GB_144_12_12} \\
\toprule
Solver & Strategy & Time (s) & Vars & Clauses & Result & Status \\
\midrule
\endhead
\midrule
\multicolumn{7}{r}{\textit{Continued on next page}} \\
\midrule
\endfoot
\bottomrule
\endlastfoot
z3py & seqcounter & 7200.000 & 1671 & 3146 & $\geq$10 & \texttt{timeout} \\
z3py & kmtotalizer & 7200.000 & 962 & 2025 & $\geq$10 & \texttt{timeout} \\
cvc5 & seqcounter & 7200.000 & 1415 & 2638 & $\geq$8 & \texttt{timeout} \\
cvc5 & kmtotalizer & 7200.000 & 979 & 2092 & $\geq$8 & \texttt{timeout} \\
minisat22 & seqcounter & 7200.000 & 3378 & 9818 & $\geq$10 & \texttt{timeout} \\
minisat22 & kmtotalizer & 7200.000 & 2669 & 8697 & $\geq$10 & \texttt{timeout} \\
glucose3 & seqcounter & 7200.000 & 3378 & 9818 & $\geq$10 & \texttt{timeout} \\
glucose3 & kmtotalizer & 7200.000 & 2669 & 8697 & $\geq$10 & \texttt{timeout} \\
\end{longtable}

\clearpage
\begin{longtable}{llrrrl p{0.22\textwidth}}
\caption{Benchmark solver results for \texttt{GB\_144\_12\_8} ($n=144$, scan $d \leq 20$, timeout 7200\,s per configuration).} \label{tab:bench-GB_144_12_8} \\
\toprule
Solver & Strategy & Time (s) & Vars & Clauses & Result & Status \\
\midrule
\endhead
\midrule
\multicolumn{7}{r}{\textit{Continued on next page}} \\
\midrule
\endfoot
\bottomrule
\endlastfoot
z3py & seqcounter & 29.836 & 1544 & 2894 & 8 & \texttt{OK} \\
z3py & kmtotalizer & 38.923 & 1010 & 2363 & 8 & \texttt{OK} \\
cvc5 & seqcounter & 18.403 & 1544 & 2894 & 8 & \texttt{OK} \\
cvc5 & kmtotalizer & 19.939 & 1010 & 2363 & 8 & \texttt{OK} \\
minisat22 & seqcounter & 6.534 & 2827 & 7870 & 8 & \texttt{OK} \\
minisat22 & kmtotalizer & 14.764 & 2293 & 7339 & 8 & \texttt{OK} \\
glucose3 & seqcounter & 10.443 & 2827 & 7870 & 8 & \texttt{OK} \\
glucose3 & kmtotalizer & 11.543 & 2293 & 7339 & 8 & \texttt{OK} \\
\end{longtable}

\clearpage
\begin{longtable}{llrrrl p{0.22\textwidth}}
\caption{Benchmark solver results for \texttt{LP\_136\_32\_4} ($n=136$, scan $d \leq 20$, timeout 7200\,s per configuration).} \label{tab:bench-LP_136_32_4} \\
\toprule
Solver & Strategy & Time (s) & Vars & Clauses & Result & Status \\
\midrule
\endhead
\midrule
\multicolumn{7}{r}{\textit{Continued on next page}} \\
\midrule
\endfoot
\bottomrule
\endlastfoot
z3py & seqcounter & 2.395 & 1000 & 1778 & 4 & \texttt{OK} \\
z3py & kmtotalizer & 2.249 & 940 & 1839 & 4 & \texttt{OK} \\
cvc5 & seqcounter & 1.403 & 1000 & 1778 & 4 & \texttt{OK} \\
cvc5 & kmtotalizer & 1.295 & 940 & 1839 & 4 & \texttt{OK} \\
minisat22 & seqcounter & 0.675 & 2400 & 7194 & 4 & \texttt{OK} \\
minisat22 & kmtotalizer & 0.474 & 2340 & 7255 & 4 & \texttt{OK} \\
glucose3 & seqcounter & 0.370 & 2400 & 7194 & 4 & \texttt{OK} \\
glucose3 & kmtotalizer & 0.320 & 2340 & 7255 & 4 & \texttt{OK} \\
\end{longtable}

\clearpage
\begin{longtable}{llrrrl p{0.22\textwidth}}
\caption{Benchmark solver results for \texttt{LP\_1768\_224\_?} ($n=1768$, scan $d \leq 20$, timeout 7200\,s per configuration).} \label{tab:bench-LP_1768_224_unknown} \\
\toprule
Solver & Strategy & Time (s) & Vars & Clauses & Result & Status \\
\midrule
\endhead
\midrule
\multicolumn{7}{r}{\textit{Continued on next page}} \\
\midrule
\endfoot
\bottomrule
\endlastfoot
z3py & seqcounter & 7200.000 & 12808 & 23186 & $\geq$5 & \texttt{timeout} \\
z3py & kmtotalizer & 7200.000 & 11796 & 22878 & $\geq$6 & \texttt{timeout} \\
cvc5 & seqcounter & 7200.000 & 12808 & 23186 & $\geq$5 & \texttt{timeout} \\
cvc5 & kmtotalizer & 7200.000 & 11796 & 22879 & $\geq$5 & \texttt{timeout} \\
minisat22 & seqcounter & 7200.000 & 63775 & 221526 & $\geq$6 & \texttt{timeout} \\
minisat22 & kmtotalizer & 7200.000 & 61515 & 221273 & $\geq$7 & \texttt{timeout} \\
glucose3 & seqcounter & 7200.000 & 63775 & 221526 & $\geq$6 & \texttt{timeout} \\
glucose3 & kmtotalizer & 7200.000 & 61515 & 221273 & $\geq$7 & \texttt{timeout} \\
\end{longtable}

\clearpage
\begin{longtable}{llrrrl p{0.22\textwidth}}
\caption{Benchmark solver results for \texttt{LP\_238\_44\_6} ($n=238$, scan $d \leq 20$, timeout 7200\,s per configuration).} \label{tab:bench-LP_238_44_6} \\
\toprule
Solver & Strategy & Time (s) & Vars & Clauses & Result & Status \\
\midrule
\endhead
\midrule
\multicolumn{7}{r}{\textit{Continued on next page}} \\
\midrule
\endfoot
\bottomrule
\endlastfoot
z3py & seqcounter & 34.821 & 2194 & 4024 & 6 & \texttt{OK} \\
z3py & kmtotalizer & 51.182 & 1685 & 3609 & 6 & \texttt{OK} \\
cvc5 & seqcounter & 18.236 & 2194 & 4024 & 6 & \texttt{OK} \\
cvc5 & kmtotalizer & 18.457 & 1685 & 3609 & 6 & \texttt{OK} \\
minisat22 & seqcounter & 15.731 & 4824 & 14246 & 6 & \texttt{OK} \\
minisat22 & kmtotalizer & 20.063 & 4315 & 13831 & 6 & \texttt{OK} \\
glucose3 & seqcounter & 14.162 & 4824 & 14246 & 6 & \texttt{OK} \\
glucose3 & kmtotalizer & 17.396 & 4315 & 13831 & 6 & \texttt{OK} \\
\end{longtable}

\clearpage
\begin{longtable}{llrrrl p{0.22\textwidth}}
\caption{Benchmark solver results for \texttt{LP\_340\_56\_8} ($n=340$, scan $d \leq 20$, timeout 7200\,s per configuration).} \label{tab:bench-LP_340_56_8} \\
\toprule
Solver & Strategy & Time (s) & Vars & Clauses & Result & Status \\
\midrule
\endhead
\midrule
\multicolumn{7}{r}{\textit{Continued on next page}} \\
\midrule
\endfoot
\bottomrule
\endlastfoot
z3py & seqcounter & 1035.526 & 3788 & 7070 & 8 & \texttt{OK} \\
z3py & kmtotalizer & 1173.064 & 2422 & 5463 & 8 & \texttt{OK} \\
cvc5 & seqcounter & 1064.009 & 3788 & 7070 & 8 & \texttt{OK} \\
cvc5 & kmtotalizer & 3894.239 & 2422 & 5463 & 8 & \texttt{OK} \\
minisat22 & seqcounter & 197.241 & 8108 & 23938 & 8 & \texttt{OK} \\
minisat22 & kmtotalizer & 435.407 & 6742 & 22331 & 8 & \texttt{OK} \\
glucose3 & seqcounter & 139.883 & 8108 & 23938 & 8 & \texttt{OK} \\
glucose3 & kmtotalizer & 407.450 & 6742 & 22331 & 8 & \texttt{OK} \\
\end{longtable}

\clearpage
\begin{longtable}{llrrrl p{0.22\textwidth}}
\caption{Benchmark solver results for \texttt{LP\_34\_20\_2} ($n=34$, scan $d \leq 20$, timeout 7200\,s per configuration).} \label{tab:bench-LP_34_20_2} \\
\toprule
Solver & Strategy & Time (s) & Vars & Clauses & Result & Status \\
\midrule
\endhead
\midrule
\multicolumn{7}{r}{\textit{Continued on next page}} \\
\midrule
\endfoot
\bottomrule
\endlastfoot
z3py & seqcounter & 0.192 & 206 & 332 & 2 & \texttt{OK} \\
z3py & kmtotalizer & 0.208 & 241 & 401 & 2 & \texttt{OK} \\
cvc5 & seqcounter & 0.137 & 206 & 332 & 2 & \texttt{OK} \\
cvc5 & kmtotalizer & 0.129 & 241 & 401 & 2 & \texttt{OK} \\
minisat22 & seqcounter & 0.008 & 496 & 1422 & 2 & \texttt{OK} \\
minisat22 & kmtotalizer & 0.009 & 531 & 1491 & 2 & \texttt{OK} \\
glucose3 & seqcounter & 0.006 & 496 & 1422 & 2 & \texttt{OK} \\
glucose3 & kmtotalizer & 0.008 & 531 & 1491 & 2 & \texttt{OK} \\
\end{longtable}

\clearpage
\begin{longtable}{llrrrl p{0.22\textwidth}}
\caption{Benchmark solver results for \texttt{LP\_442\_68\_10} ($n=442$, scan $d \leq 20$, timeout 7200\,s per configuration).} \label{tab:bench-LP_442_68_10} \\
\toprule
Solver & Strategy & Time (s) & Vars & Clauses & Result & Status \\
\midrule
\endhead
\midrule
\multicolumn{7}{r}{\textit{Continued on next page}} \\
\midrule
\endfoot
\bottomrule
\endlastfoot
z3py & seqcounter & 7200.000 & 4507 & 8372 & $\geq$8 & \texttt{timeout} \\
z3py & kmtotalizer & 7200.000 & 3097 & 6612 & $\geq$8 & \texttt{timeout} \\
cvc5 & seqcounter & 7200.000 & 4507 & 8372 & $\geq$8 & \texttt{timeout} \\
cvc5 & kmtotalizer & 7200.000 & 3097 & 6613 & $\geq$7 & \texttt{timeout} \\
minisat22 & seqcounter & 7200.000 & 10964 & 32818 & $\geq$9 & \texttt{timeout} \\
minisat22 & kmtotalizer & 7200.000 & 9190 & 30765 & $\geq$9 & \texttt{timeout} \\
glucose3 & seqcounter & 7200.000 & 10964 & 32818 & $\geq$9 & \texttt{timeout} \\
glucose3 & kmtotalizer & 7200.000 & 9190 & 30765 & $\geq$9 & \texttt{timeout} \\
\end{longtable}

\clearpage
\begin{longtable}{llrrrl p{0.22\textwidth}}
\caption{Benchmark solver results for \texttt{LP\_544\_80\_12} ($n=544$, scan $d \leq 20$, timeout 7200\,s per configuration).} \label{tab:bench-LP_544_80_12} \\
\toprule
Solver & Strategy & Time (s) & Vars & Clauses & Result & Status \\
\midrule
\endhead
\midrule
\multicolumn{7}{r}{\textit{Continued on next page}} \\
\midrule
\endfoot
\bottomrule
\endlastfoot
z3py & seqcounter & 7200.000 & 5020 & 9262 & $\geq$7 & \texttt{timeout} \\
z3py & kmtotalizer & 7200.000 & 3803 & 8193 & $\geq$7 & \texttt{timeout} \\
cvc5 & seqcounter & 7200.000 & 5020 & 9262 & $\geq$7 & \texttt{timeout} \\
cvc5 & kmtotalizer & 7200.000 & 3803 & 8193 & $\geq$7 & \texttt{timeout} \\
minisat22 & seqcounter & 7200.000 & 13856 & 41842 & $\geq$9 & \texttt{timeout} \\
minisat22 & kmtotalizer & 7200.000 & 11579 & 38656 & $\geq$8 & \texttt{timeout} \\
glucose3 & seqcounter & 7200.000 & 13856 & 41842 & $\geq$9 & \texttt{timeout} \\
glucose3 & kmtotalizer & 7200.000 & 11579 & 38656 & $\geq$8 & \texttt{timeout} \\
\end{longtable}

\clearpage
\begin{longtable}{llrrrl p{0.22\textwidth}}
\caption{Benchmark solver results for \texttt{LP\_714\_100\_?} ($n=714$, scan $d \leq 20$, timeout 7200\,s per configuration).} \label{tab:bench-LP_714_100_unknown} \\
\toprule
Solver & Strategy & Time (s) & Vars & Clauses & Result & Status \\
\midrule
\endhead
\midrule
\multicolumn{7}{r}{\textit{Continued on next page}} \\
\midrule
\endfoot
\bottomrule
\endlastfoot
z3py & seqcounter & 7200.000 & 6590 & 12172 & $\geq$7 & \texttt{timeout} \\
z3py & kmtotalizer & 7200.000 & 4937 & 10377 & $\geq$7 & \texttt{timeout} \\
cvc5 & seqcounter & 7200.000 & 6590 & 12172 & $\geq$7 & \texttt{timeout} \\
cvc5 & kmtotalizer & 7200.000 & 4937 & 10377 & $\geq$7 & \texttt{timeout} \\
minisat22 & seqcounter & 7200.000 & 19548 & 60370 & $\geq$9 & \texttt{timeout} \\
minisat22 & kmtotalizer & 7200.000 & 16495 & 55778 & $\geq$8 & \texttt{timeout} \\
glucose3 & seqcounter & 7200.000 & 19548 & 60370 & $\geq$9 & \texttt{timeout} \\
glucose3 & kmtotalizer & 7200.000 & 16495 & 55778 & $\geq$8 & \texttt{timeout} \\
\end{longtable}

\clearpage
\begin{longtable}{llrrrl p{0.22\textwidth}}
\caption{Benchmark solver results for \texttt{TN\_108\_2\_12} ($n=108$, scan $d \leq 20$, timeout 7200\,s per configuration).} \label{tab:bench-TN_108_2_12} \\
\toprule
Solver & Strategy & Time (s) & Vars & Clauses & Result & Status \\
\midrule
\endhead
\midrule
\multicolumn{7}{r}{\textit{Continued on next page}} \\
\midrule
\endfoot
\bottomrule
\endlastfoot
z3py & seqcounter & 236.588 & 1480 & 2826 & 12 & \texttt{OK} \\
z3py & kmtotalizer & 396.838 & 729 & 1639 & 12 & \texttt{OK} \\
cvc5 & seqcounter & 7200.000 & 1308 & 2486 & $\geq$11 & \texttt{timeout} \\
cvc5 & kmtotalizer & 7200.000 & 714 & 1512 & $\geq$11 & \texttt{timeout} \\
minisat22 & seqcounter & 190.268 & 2394 & 6370 & 12 & \texttt{OK} \\
minisat22 & kmtotalizer & 461.414 & 1643 & 5183 & 12 & \texttt{OK} \\
glucose3 & seqcounter & 129.319 & 2394 & 6370 & 12 & \texttt{OK} \\
glucose3 & kmtotalizer & 193.134 & 1643 & 5183 & 12 & \texttt{OK} \\
\end{longtable}

\clearpage
\begin{longtable}{llrrrl p{0.22\textwidth}}
\caption{Benchmark solver results for \texttt{TN\_144\_2\_13} ($n=144$, scan $d \leq 20$, timeout 7200\,s per configuration).} \label{tab:bench-TN_144_2_13} \\
\toprule
Solver & Strategy & Time (s) & Vars & Clauses & Result & Status \\
\midrule
\endhead
\midrule
\multicolumn{7}{r}{\textit{Continued on next page}} \\
\midrule
\endfoot
\bottomrule
\endlastfoot
z3py & seqcounter & 7200.000 & 1899 & 3630 & $\geq$12 & \texttt{timeout} \\
z3py & kmtotalizer & 7200.000 & 942 & 2016 & $\geq$11 & \texttt{timeout} \\
cvc5 & seqcounter & 7200.000 & 1651 & 3138 & $\geq$10 & \texttt{timeout} \\
cvc5 & kmtotalizer & 7200.000 & 990 & 2355 & $\geq$9 & \texttt{timeout} \\
minisat22 & seqcounter & 7200.000 & 3145 & 8466 & $\geq$12 & \texttt{timeout} \\
minisat22 & kmtotalizer & 7200.000 & 2188 & 6851 & $\geq$12 & \texttt{timeout} \\
glucose3 & seqcounter & 7200.000 & 3145 & 8466 & $\geq$12 & \texttt{timeout} \\
glucose3 & kmtotalizer & 7200.000 & 2188 & 6851 & $\geq$12 & \texttt{timeout} \\
\end{longtable}

\clearpage
\begin{longtable}{llrrrl p{0.22\textwidth}}
\caption{Benchmark solver results for \texttt{TN\_200\_10\_10} ($n=200$, scan $d \leq 20$, timeout 7200\,s per configuration).} \label{tab:bench-TN_200_10_10} \\
\toprule
Solver & Strategy & Time (s) & Vars & Clauses & Result & Status \\
\midrule
\endhead
\midrule
\multicolumn{7}{r}{\textit{Continued on next page}} \\
\midrule
\endfoot
\bottomrule
\endlastfoot
z3py & seqcounter & 371.667 & 2520 & 4794 & 10 & \texttt{OK} \\
z3py & kmtotalizer & 913.709 & 1366 & 2904 & 10 & \texttt{OK} \\
cvc5 & seqcounter & 331.767 & 2520 & 4794 & 10 & \texttt{OK} \\
cvc5 & kmtotalizer & 921.306 & 1366 & 2904 & 10 & \texttt{OK} \\
minisat22 & seqcounter & 264.878 & 4492 & 12470 & 10 & \texttt{OK} \\
minisat22 & kmtotalizer & 561.131 & 3338 & 10580 & 10 & \texttt{OK} \\
glucose3 & seqcounter & 284.763 & 4492 & 12470 & 10 & \texttt{OK} \\
glucose3 & kmtotalizer & 226.393 & 3338 & 10580 & 10 & \texttt{OK} \\
\end{longtable}

\clearpage
\begin{longtable}{llrrrl p{0.22\textwidth}}
\caption{Benchmark solver results for \texttt{TN\_250\_10\_15} ($n=250$, scan $d \leq 20$, timeout 7200\,s per configuration).} \label{tab:bench-TN_250_10_15} \\
\toprule
Solver & Strategy & Time (s) & Vars & Clauses & Result & Status \\
\midrule
\endhead
\midrule
\multicolumn{7}{r}{\textit{Continued on next page}} \\
\midrule
\endfoot
\bottomrule
\endlastfoot
z3py & seqcounter & 7200.000 & 3399 & 6498 & $\geq$12 & \texttt{timeout} \\
z3py & kmtotalizer & 7200.000 & 1616 & 3402 & $\geq$11 & \texttt{timeout} \\
cvc5 & seqcounter & 7200.000 & 3170 & 6042 & $\geq$11 & \texttt{timeout} \\
cvc5 & kmtotalizer & 7200.000 & 1616 & 3403 & $\geq$10 & \texttt{timeout} \\
minisat22 & seqcounter & 7200.000 & 5893 & 16214 & $\geq$12 & \texttt{timeout} \\
minisat22 & kmtotalizer & 7200.000 & 4110 & 13118 & $\geq$11 & \texttt{timeout} \\
glucose3 & seqcounter & 7200.000 & 5893 & 16214 & $\geq$12 & \texttt{timeout} \\
glucose3 & kmtotalizer & 7200.000 & 4110 & 13117 & $\geq$12 & \texttt{timeout} \\
\end{longtable}

\clearpage
\begin{longtable}{llrrrl p{0.22\textwidth}}
\caption{Benchmark solver results for \texttt{TN\_360\_4\_?} ($n=360$, scan $d \leq 20$, timeout 7200\,s per configuration).} \label{tab:bench-TN_360_4_unknown} \\
\toprule
Solver & Strategy & Time (s) & Vars & Clauses & Result & Status \\
\midrule
\endhead
\midrule
\multicolumn{7}{r}{\textit{Continued on next page}} \\
\midrule
\endfoot
\bottomrule
\endlastfoot
z3py & seqcounter & 7200.000 & 4247 & 8110 & $\geq$10 & \texttt{timeout} \\
z3py & kmtotalizer & 7200.000 & 2434 & 5189 & $\geq$10 & \texttt{timeout} \\
cvc5 & seqcounter & 7200.000 & 3559 & 6738 & $\geq$8 & \texttt{timeout} \\
cvc5 & kmtotalizer & 7200.000 & 2395 & 5177 & $\geq$7 & \texttt{timeout} \\
minisat22 & seqcounter & 7200.000 & 7934 & 21806 & $\geq$11 & \texttt{timeout} \\
minisat22 & kmtotalizer & 7200.000 & 5780 & 18205 & $\geq$10 & \texttt{timeout} \\
glucose3 & seqcounter & 7200.000 & 7934 & 21806 & $\geq$11 & \texttt{timeout} \\
glucose3 & kmtotalizer & 7200.000 & 5780 & 18204 & $\geq$11 & \texttt{timeout} \\
\end{longtable}

\clearpage
\begin{longtable}{llrrrl p{0.22\textwidth}}
\caption{Benchmark solver results for \texttt{TN\_36\_8\_3} ($n=36$, scan $d \leq 20$, timeout 7200\,s per configuration).} \label{tab:bench-TN_36_8_3} \\
\toprule
Solver & Strategy & Time (s) & Vars & Clauses & Result & Status \\
\midrule
\endhead
\midrule
\multicolumn{7}{r}{\textit{Continued on next page}} \\
\midrule
\endfoot
\bottomrule
\endlastfoot
z3py & seqcounter & 0.237 & 223 & 386 & 3 & \texttt{OK} \\
z3py & kmtotalizer & 0.252 & 229 & 396 & 3 & \texttt{OK} \\
cvc5 & seqcounter & 0.150 & 223 & 386 & 3 & \texttt{OK} \\
cvc5 & kmtotalizer & 0.176 & 229 & 396 & 3 & \texttt{OK} \\
minisat22 & seqcounter & 0.011 & 475 & 1346 & 3 & \texttt{OK} \\
minisat22 & kmtotalizer & 0.010 & 481 & 1356 & 3 & \texttt{OK} \\
glucose3 & seqcounter & 0.010 & 475 & 1346 & 3 & \texttt{OK} \\
glucose3 & kmtotalizer & 0.010 & 481 & 1356 & 3 & \texttt{OK} \\
\end{longtable}

\clearpage
\begin{longtable}{llrrrl p{0.22\textwidth}}
\caption{Benchmark solver results for \texttt{TN\_36\_8\_4} ($n=36$, scan $d \leq 20$, timeout 7200\,s per configuration).} \label{tab:bench-TN_36_8_4} \\
\toprule
Solver & Strategy & Time (s) & Vars & Clauses & Result & Status \\
\midrule
\endhead
\midrule
\multicolumn{7}{r}{\textit{Continued on next page}} \\
\midrule
\endfoot
\bottomrule
\endlastfoot
z3py & seqcounter & 0.379 & 252 & 446 & 4 & \texttt{OK} \\
z3py & kmtotalizer & 0.339 & 244 & 471 & 4 & \texttt{OK} \\
cvc5 & seqcounter & 0.233 & 252 & 446 & 4 & \texttt{OK} \\
cvc5 & kmtotalizer & 0.197 & 244 & 471 & 4 & \texttt{OK} \\
minisat22 & seqcounter & 0.024 & 604 & 1802 & 4 & \texttt{OK} \\
minisat22 & kmtotalizer & 0.028 & 596 & 1827 & 4 & \texttt{OK} \\
glucose3 & seqcounter & 0.023 & 604 & 1802 & 4 & \texttt{OK} \\
glucose3 & kmtotalizer & 0.024 & 596 & 1827 & 4 & \texttt{OK} \\
\end{longtable}

\clearpage
\begin{longtable}{llrrrl p{0.22\textwidth}}
\caption{Benchmark solver results for \texttt{TN\_54\_11\_4} ($n=54$, scan $d \leq 20$, timeout 7200\,s per configuration).} \label{tab:bench-TN_54_11_4} \\
\toprule
Solver & Strategy & Time (s) & Vars & Clauses & Result & Status \\
\midrule
\endhead
\midrule
\multicolumn{7}{r}{\textit{Continued on next page}} \\
\midrule
\endfoot
\bottomrule
\endlastfoot
z3py & seqcounter & 0.421 & 384 & 680 & 4 & \texttt{OK} \\
z3py & kmtotalizer & 0.457 & 364 & 687 & 4 & \texttt{OK} \\
cvc5 & seqcounter & 0.265 & 384 & 680 & 4 & \texttt{OK} \\
cvc5 & kmtotalizer & 0.307 & 364 & 687 & 4 & \texttt{OK} \\
minisat22 & seqcounter & 0.026 & 776 & 2178 & 4 & \texttt{OK} \\
minisat22 & kmtotalizer & 0.032 & 756 & 2185 & 4 & \texttt{OK} \\
glucose3 & seqcounter & 0.031 & 776 & 2178 & 4 & \texttt{OK} \\
glucose3 & kmtotalizer & 0.029 & 756 & 2185 & 4 & \texttt{OK} \\
\end{longtable}

\clearpage
\begin{longtable}{llrrrl p{0.22\textwidth}}
\caption{Benchmark solver results for \texttt{TN\_72\_14\_4} ($n=72$, scan $d \leq 20$, timeout 7200\,s per configuration).} \label{tab:bench-TN_72_14_4} \\
\toprule
Solver & Strategy & Time (s) & Vars & Clauses & Result & Status \\
\midrule
\endhead
\midrule
\multicolumn{7}{r}{\textit{Continued on next page}} \\
\midrule
\endfoot
\bottomrule
\endlastfoot
z3py & seqcounter & 0.632 & 516 & 918 & 4 & \texttt{OK} \\
z3py & kmtotalizer & 0.716 & 488 & 947 & 4 & \texttt{OK} \\
cvc5 & seqcounter & 0.374 & 516 & 918 & 4 & \texttt{OK} \\
cvc5 & kmtotalizer & 0.387 & 488 & 947 & 4 & \texttt{OK} \\
minisat22 & seqcounter & 0.070 & 1024 & 2858 & 4 & \texttt{OK} \\
minisat22 & kmtotalizer & 0.079 & 996 & 2887 & 4 & \texttt{OK} \\
glucose3 & seqcounter & 0.063 & 1024 & 2858 & 4 & \texttt{OK} \\
glucose3 & kmtotalizer & 0.057 & 996 & 2887 & 4 & \texttt{OK} \\
\end{longtable}

\clearpage
\begin{longtable}{llrrrl p{0.22\textwidth}}
\caption{Benchmark solver results for \texttt{TN\_72\_8\_8} ($n=72$, scan $d \leq 20$, timeout 7200\,s per configuration).} \label{tab:bench-TN_72_8_8} \\
\toprule
Solver & Strategy & Time (s) & Vars & Clauses & Result & Status \\
\midrule
\endhead
\midrule
\multicolumn{7}{r}{\textit{Continued on next page}} \\
\midrule
\endfoot
\bottomrule
\endlastfoot
z3py & seqcounter & 6.438 & 744 & 1386 & 8 & \texttt{OK} \\
z3py & kmtotalizer & 10.304 & 506 & 1167 & 8 & \texttt{OK} \\
cvc5 & seqcounter & 4.639 & 744 & 1386 & 8 & \texttt{OK} \\
cvc5 & kmtotalizer & 8.208 & 506 & 1167 & 8 & \texttt{OK} \\
minisat22 & seqcounter & 2.889 & 1464 & 4178 & 8 & \texttt{OK} \\
minisat22 & kmtotalizer & 6.285 & 1226 & 3959 & 8 & \texttt{OK} \\
glucose3 & seqcounter & 3.440 & 1464 & 4178 & 8 & \texttt{OK} \\
glucose3 & kmtotalizer & 4.379 & 1226 & 3959 & 8 & \texttt{OK} \\
\end{longtable}

\clearpage
\begin{figure*}[p]
    \centering

    \includegraphics[
        width=1\textwidth
    ]{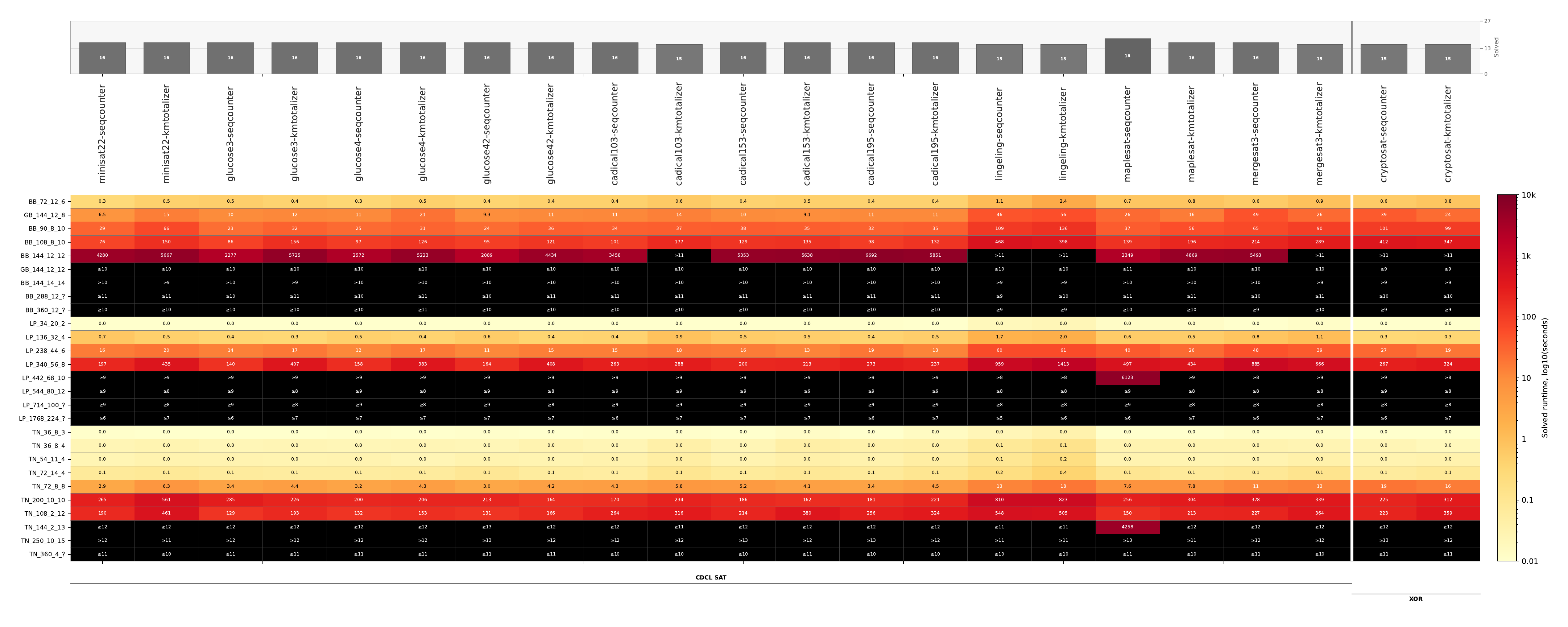}\\

    \includegraphics[
        width=1\textwidth
    ]{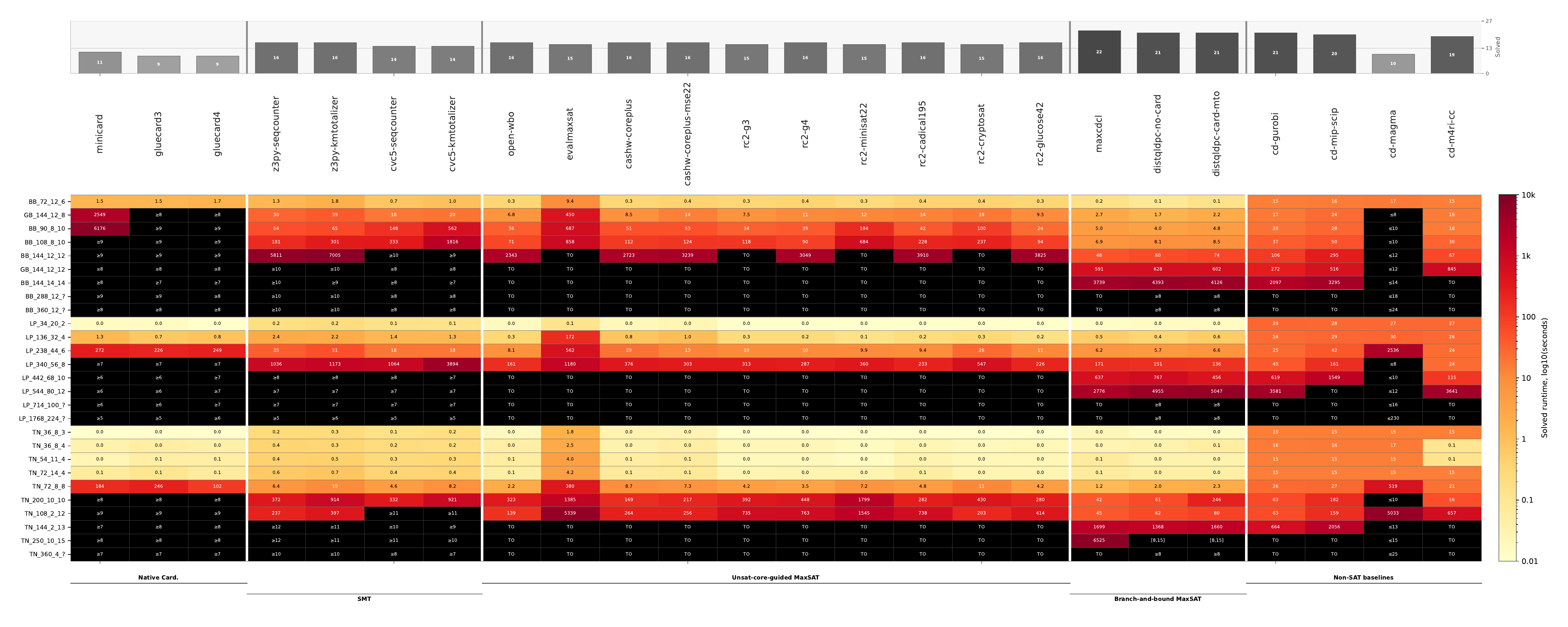}
    \caption{
    Full solver-behavior heatmap. We separate the solver family groups with white bars and place the family names below the heatmap. One can clearly see that the differences between groups are much larger than the differences within groups.
    Light colors indicate faster solved runtimes, while black cells denote timeout runs.
    Cell: run time in seconds. Black Cell: timeout with obtained distance upper/lower bounds. Bar chart: number of solved cases within 2 hours.
    }
    \label{fig:full-heatmap}
\end{figure*}
\end{document}